\documentclass[aps,preprint]{revtex4}%
\usepackage{amsfonts}
\usepackage{amsmath}
\usepackage{amssymb}
\usepackage{graphicx}%
\begin{document}
\title{Model Predictions for the Phenomenology of Isoscalar Heavy Baryons }
\author{Z. Aziza Baccouche and Thomas D. Cohen}
\affiliation{Department of Physics, University of Maryland, College Park, MD 20742-4111.}

\begin{abstract}
We study certain physical observables of isoscalar heavy baryons using
potential models based on hadronic degrees of freedom. The goal is to compare
these results with those from an effective theory obtained in a combined heavy
quark and large $N_{c}$ expansion and thereby test the convergence of the
effective theory. We study the excitation energy of $\Lambda_{b}$ and
semileptonic decay form factors. The models tend to confirm the usefulness of
the effective theory for most observables.

\end{abstract}
\maketitle

\section{Introduction}

\label{Intro}
\def\baselinestretch{1.0}\small\normalsize
In a series of previous papers \cite{hb0,hb1,hb2,hb3,hb4}, the phenomenology
of isoscalar heavy baryons (where by heavy baryon we mean one containing a
charm or bottom quark) and their excited states were studied in a combined
heavy quark and large $N_{c}$ expansion where the natural counting parameter
is in terms of $\lambda^{\frac{1}{2}}$ (where $\lambda=\frac{1}{N_{c}}%
,\frac{\Lambda}{m_{Q}}$ with $N_{c}$ denoting the number of QCD colors,
$m_{Q}$ represents the heavy quark mass, and $\Lambda$ is a typical hadronic
scale). In this combined expansion, the ratio $N_{c}\frac{\Lambda_{had}}%
{m_{Q}}$ is arbitrary. As $\lambda\rightarrow0$ in the combined limit,
QCD\ acquires a dynamical symmetry---a contracted $O(8)$ symmetry, which
connects orbitally excited states of the heavy baryon to its ground state
\cite{hb1}. \ This contracted $O(8)$ symmetry group contains as a subgroup the
generating algebra of a simple harmonic oscillator. In ref. \cite{hb4},
predictions at next-to-leading order (NLO) in this combined expansion have
been made for the energy levels, semi-leptonic decays, and electromagnetic
decay rates of $\Lambda_{c}$ and $\Lambda_{b}$ baryons. Although the effective
theory with its approximate symmetry can be derived directly from QCD
\cite{hb0}, these heavy baryons can be described in terms of models based on a
bound state picture which describes the heavy baryon as a bound state of an
ordinary baryon with a heavy meson
\cite{bst1,bst2,bst3,bst4,bst5,bst6,bst7,bst8,bst9}. These bound state models
exhibit the same QCD contracted $O(8)$ symmetry with similar counting rules
for the effective operators.

While the predictions in ref. \cite{hb4} based on the effective theory are
model independent, they do depend on the convergence of the effective
expansion. In this paper, we explore the issue of convergence of the expansion
by comparing the theoretical predictions based on the expansion derived in
ref. \cite{hb3} with predictions based on \textquotedblleft
realistic\textquotedblright\ models. Our goal is to see whether for these
models the contribution of higher order terms in the effective expansion are
large enough to render the expansion useless or, conversely, whether the
expansion proves to be relatively accurate. To do this, we study various
potential models which are based on the bound state picture of heavy baryons.
We will choose \textquotedblleft reasonable\textquotedblright\ potential
models with scales typical of hadronic physics. These models are supposed to
describe the states of heavy baryons below threshold between the dissociation
into an ordinary light baryon and a heavy meson. The effective theory ought to
well describe the low-lying states below threshold provided the expansion is
convergent. We will test the model predictions for the physical observables of
low--lying $\Lambda_{c}$ and $\Lambda_{b}$ baryons against the effective
theory at NLO as a way to get a qualitative sense on how convergent the
expansion is expected to be in nature. Of course, these models are
\textit{not} nature and thus there is no guarantee that effective convergence
in any given model insures that in nature the expansion ought to converge.
However, if the validity of the expansion is robust in the sense that the
effects of the higher order terms are small for a wide variety of sensible
models, it is plausible that the expansion will be useful for the real world.

Heavy baryons can be envisioned in terms of two complementary bound state
pictures; one picture describes the heavy baryon as the bound state of a heavy
meson (containing the heavy quark) coupled to an ordinary light baryon. Models
of this sort based on the Skyrme model were commonly studied in the early
1990s \cite{bst1,bst2,bst3,bst4,bst5}. We will refer to this particular
picture as the bound state picture. Another way to describe the heavy baryon
is in terms of a single heavy quark coupled to the \textit{brown-muck---}a
technical term for the light degrees of freedom which includes the light
quarks and gluons. The connection of these pictures with QCD\ is
straightforward. The key point is that the motion between the heavy quark and
the brown-muck or, equivalently, between the heavy meson and the ordinary
light baryon is collective. This motion is characteristically low energy with
a characteristic excitation energy of order $\lambda^{\frac{1}{2}}$
\cite{hb3}, while the internal excitations of the brown-muck (or equivalently
of the ordinary light baryon) are of order $\lambda^{0}$. This scale
separation is what permits the development of an effective theory.

In this paper, we will work with the bound state picture which involves the
bound state of a heavy meson and light baryon. This picture has the advantage
that the heavy meson and light baryon masses are well defined and
experimentally accessible in an unambiguous way. This is not the case,
however, for the heavy quark and brown-muck masses. In the combined limit, the
heavy meson mass and the light baryon mass both scale as $\lambda^{-1}$ since
the light baryon mass $m_{bar}\sim N_{c}$ (because of the large $N_{c}$ limit
\cite{LN1,LN2}) and the heavy meson mass $m_{H}\sim m_{Q}$ (because of \ the
heavy quark limit \cite{HQ1,HQ2,HQ3,HQ4,HQ5,HQ6}). Because both of these
masses are heavy, the collective wave function describing the low-energy
excitations is localized near the bottom of the effective potential and does
not spread appreciably. The collective motion describing the low-lying states
of heavy baryons is approximately harmonic since all non-singular potentials
look harmonic at the bottom of any potential well. The excitation energy of
these low-lying states at leading order (LO) is given by,
\begin{equation}
\omega=\sqrt{\frac{\kappa}{\mu_{Q}}},
\end{equation}
where $\kappa$ is the effective spring constant. By general large $N_{c}$
counting rules \cite{LN1,LN2}, $\kappa$ scales as $N_{c}^{0}\thicksim
\lambda^{0}$, and $\mu_{Q}=\frac{m_{bar}m_{H}}{m_{bar}+m_{H}}\thicksim
\lambda^{-1}$ is the reduced mass of the bound state picture where $m_{bar}$
is the light baryon mass and $m_{H}$ denotes the heavy meson mass. As
$\lambda\rightarrow0$, the whole tower of harmonically excited states becomes
degenerate with the ground state---a clear signature of an emergent symmetry.
This symmetry is a contracted $O(8)$ symmetry which connects orbitally excited
states of the heavy baryon to its ground state via raising and lowering
operators. For small $\lambda$, the low-lying collective states of the heavy
baryon are well described in terms of harmonic oscillator wave functions. This
approximation, however, clearly breaks down near threshold where the states
become anharmonic.

A generic potential model can have an arbitrary number of free parameters.
However, since we are using the potential models to test our effective theory
at NLO (at this order the physical observables of $\Lambda_{c}$ and
$\Lambda_{b}$ baryons are described in terms of two relevant parameters), it
is convenient to study potential models which also have two free parameters.
We can use this freedom to fix the observables to experiment. We can then make
predictions for other observables, and further test how well the expansion
works. Accordingly, we chose potential models with two parameters---the
potential depth and a length parameter. For simplicity, we begin with a study
of single-channel potentials. Of course, any model based on physical hadrons
couples the light baryon to a physical heavy meson. The physical heavy meson
states are denoted by H and H$^{\ast}$(where H is a heavy meson with spin
$J=0,$ and H$^{\ast}$ is a heavy meson with $J=1$.) Recall, however, that in
the heavy quark limit H and H$^{\ast}$are degenerate and, from the point of
view of the collective dynamics, one can construct a simple model where we
neglect the splitting and treat the two states as a generic heavy meson. In
the charmed sector, for example, the two mesons are the D and D$^{\ast}$mesons
which, in these simple models, can be treated as a single meson with a mass
taken to be the spin-averaged mass of the two states. Note that this
single-channel model is consistent with the $\lambda^{\frac{1}{2}}$ expansion
at NLO since the mass splitting between the heavy mesons contributes only at NNLO.

By using potential models, we can calculate a number of observables for
$\Lambda_{Q}$ (whereby $Q$ we mean a charm or bottom quark) baryons. We first
compute the spectroscopic observables for these baryons. We should note
however that, since the effective theory clearly breaks down beyond the first
doublet of excited states, we will restrict our attention to the excitation
energy of the first excited state. We also study semi-leptonic decays of
$\Lambda_{c}$ and $\Lambda_{b}$ baryons. These decays involve transitions
between the ground state of $\Lambda_{b}$ and the ground state of $\Lambda
_{c}$ as well as transitions between the ground state of $\Lambda_{b}$ and the
doublet of first excited states of $\Lambda_{c}$. These observables have the
virtue that, at NLO in the $\lambda^{\frac{1}{2}}$ expansion, the currents
associated with these observables are determined entirely by the symmetries of
the effective theory with no free parameters \cite{hb2}. Thus, these
observables can be predicted at NLO without additional fitting. Of course, in
a general potential model, there is no reason why these currents need to be
exactly of these forms. However, we can take them to be of these simple forms
in order to reduce the number of free parameters. With these currents, we can
compare the model predictions for these observables to the predictions based
on the effective theory at NLO in the expansion. Since we have made a number
of arbitrary assumptions about the form of the potentials and choice of
currents, this comparison cannot be taken as definitive in determining the
degree of convergence of the expansion for the various observables. It should,
however, provide a qualitative test of convergence.

We can also generalize the potential models to include coupled channels where
the heavy meson states H and H$^{\ast}$ are treated as two distinct states
coupled separately to the light baryon. This occurs only at NNLO in the
$\lambda$ expansion. Accordingly, the mass splitting effects of these meson
states are beyond the order to which we have worked in the expansion. Thus,
these models are a useful place to look for effects which may spoil the
expansion. While these coupled-channel potential models produce more states
than the single-channel potentials, these models should agree with the
expansion (and hence with the single-channel models) for low-lying states in
the limit where $\lambda\rightarrow0$. Moreover, we find that there is no way
to match these theories unless we simultaneously include at least one
excitation of the light baryon (or equivalently the brown-muck). In doing
this, we effectively obtain, in the $\lambda\rightarrow0$ limit, two decoupled
copies of the collective motion: one based on the ground state of the light
baryon oscillating against the generic heavy meson, and one on its excited
state. Indeed, in the general derivation of the effective theory, it was
pointed out that, since the intrinsic structure of the brown-muck is
associated with qualitatively faster dynamics than the collective motion, the
collective motion can be built on either the ground state of the brown-muck or
its excited states. In fact, as was shown in ref. \cite{hb2}, the collective
motion is essentially the same for both cases. As $\lambda$ deviates from
zero, these states mix due to the splitting of the heavy mesons.

The construction of such coupled-channel potentials is subtle. The collective
dynamics clearly favor s-waves for the ground state. Thus, the parity of the
ground state collective wave function is even. The parity of the ground state
heavy baryon is also even. Thus, the product of the intrinsic parities of the
light baryon and the heavy meson in the model must also be even. The H and
H$^{\ast}$ mesons both have odd intrinsic parity. Thus, in order to obtain
states consistent with the parity of $\Lambda_{Q}$ baryons, these heavy mesons
have to be coupled to a light baryon with odd intrinsic parity. Accordingly,
in such models, we will not use the nucleon since it has a positive intrinsic
parity. Instead, we consider an excited state of the nucleon with
negative---odd---parity. For concreteness, we will use the N$^{\ast}$(1535)
nucleon resonance state as our low-lying light baryon state and the
N$^{\ast\ast}$(2090) nucleon resonance state as our excited light baryon. We
note in passing that previous model treatments based on the bound state
picture have used Skyrmions for the light baryon
\cite{bst1,bst2,bst3,bst4,bst5,bst6,bst7,bst8,bst9}; in these models, the
Skyrmion implicitly includes baryons of both parities and excitations. Thus,
the treatment here is in some ways analogous to previous work. The present
models, however, allow for detailed dynamical calculations.

Again, in principle, there are an infinite number of arbitrary parameters that
arise in building the most general coupled-channel potential. Here we will
pick arbitrary forms with two free parameters. Our strategy will be as
follows: we start with a single-channel potential with two free parameters. We
consider the limit of this potential where $\lambda\rightarrow0$. In this
limit, the form of the coupling of the two channels is fixed. We keep fixed
the relative strengths of the two channels as found in this limit and then
depart from the $\lambda=0$ limit by using the full potential form and
including the mass splitting of the heavy mesons. This strategy guarantees the
correct behavior as the symmetry limit is approached. Again, this is not the
most general approach one can take, but it should provide insights about
convergence. One important limitation to this approach is that we do not build
in any spin-orbit interactions. Thus, in these models $L$ is a good quantum
number. Moreover, the absence of spin-orbit interactions means that the
doublets of states for $L\neq0$ are all unsplit. Of course, at the expense of
an additional parameter, we could build in the spin-orbit interaction.
However, doing this will not provide us with any additional qualitative
insights about the convergence of the expansion, and thus we have chosen not
to do this.

As will be shown in this paper, the model-independent predictions based on the
effective expansion work rather well for these observables. Naively, since
$\lambda\sim\frac{1}{N_{c}}$ and the expansion is in $\lambda^{\frac{1}{2}}$,
one would expect typical errors at NLO to be of order $\frac{1}{3}%
\thickapprox33\%$. In fact, we see that most of the model-independent
predictions based on the expansion \cite{hb4} seem to work at least this well
for all of the models studied. This gives us some confidence that, as
experimental measurements of these quantities become available, the
model-independent predictions of ref.~\cite{hb4} may give us a reliable guide.
The exception to this general trend is predictions of the electromagnetic
decay widths. The reason for the relatively poor behavior of these observables
is largely due to the fact that charge assignments for quarks in the large
$N_{c}$ limit differ substantially from those in the real world \cite{hb2}.
Thus, these observables are known to have very large $\frac{1}{N_{c}}$
corrections and are not expected to give accurate predictions. Accordingly,
calculations of these observables will not be reported here.

We will also see, however, that there are large differences between the single
and coupled-channel model predictions for the semi-leptonic decays. As these
differences come entirely from effects at NNLO and beyond in the effective
theory, these effects would ordinarily be small provided the expansion were
well converged at NNLO. We suspect that the success of the NLO
model-independent predictions for the semi-leptonic decays, despite the clear
evidence of large NNLO effects, is due to correlations in the combinations of
observables in our model-independent predictions which render them relatively
insensitive to higher order effects.

This paper is organized as follows. In the next section, we provide a brief
review of the model-independent predictions at NLO in the expansion. Next, we
discuss the results based on a simple single-channel model. Following this, we
discuss how to derive coupled-channel potentials consistent with our
expansion. Finally, we compare the results based on the potential models to
the model-independent predictions based on the effective theory at NLO.

\section{Model-Independent Predictions Based on the Effective Expansion}

\label{ModIndPred}In this section, the model-independent predictions of ref.
\cite{hb3,hb4} will briefly be reviewed. The effective theory is expressed in
terms of collective operators. These operators are defined unambiguously in
QCD up to order $\lambda$ and describe the position and momentum of the
brown-muck relative to the heavy quark as well as the total position and
momentum of the system. At NLO, the relative variables are completely
equivalent to the position and momentum of an ordinary light baryon relative
to a heavy meson. The derivation is detailed in ref.~\cite{hb1}. Apart from
the general QCD large $N_{c}$ and heavy quark counting rules, there is an
additional symmetry issue. If the minimum of the effective potential is at
zero separation (the symmetric case) which we assume here, then the following
counting rules apply:%

\begin{align}
(\overrightarrow{x},\,\overrightarrow{X})\,  &  \sim\,\lambda^{\frac{1}{4}%
}\,,\label{CRules}\\
(\overrightarrow{p},\,\,\overrightarrow{P})\,  &  \sim\,\lambda^{-\frac{1}{4}%
}\,, \label{CR}%
\end{align}
where $\overrightarrow{x}$ and $\overrightarrow{p}$ are the relative position
and momentum, and $\overrightarrow{X}$ and $\overrightarrow{P}$ are the total
position and momentum of the heavy quark and brown-muck \cite{hb1}.

Including terms of $\mathcal{O}(\lambda)$, the effective Hamiltonian based on
the counting rules of eq. (\ref{CRules}) was derived in ref. \cite{hb1},
\begin{equation}%
\begin{array}
[c]{ccccccccccccccc}%
\mathcal{H}_{eff} & = & (m_{H}\,+\,m_{bar}) & \,+\, & {c}_{0} & \,+\, &
\left(  \frac{P^{2}}{2(m_{bar}+m_{H})}+\frac{p^{2}}{2\mu_{Q}}\,+\,\frac{1}%
{2}\kappa x^{2}\right)  & \,+\, & \frac{1}{4!}\alpha x^{4} & \,+\, &
\mathcal{O}(\lambda^{\frac{3}{2}})\,, &  &  &  & \\
&  & \parallel &  & \parallel &  & \parallel &  & \parallel &  &  &  &  &  &
\\
&  & \mathcal{H}_{\lambda^{-1}} &  & \mathcal{H}_{\lambda^{0}} &  &
\mathcal{H}_{\lambda^{\frac{1}{2}}} &  & \mathcal{H}_{\lambda^{1}} &  &  &  &
&  &
\end{array}
\label{EffHam}%
\end{equation}
where $c_{0}$ is the potential depth relative to the dissociation threshold,
$m_{bar}$ is the light baryon mass, $m_{H}$ is the spin-averaged heavy meson
mass, and $\mu_{Q}=\frac{m_{bar}m_{H}}{m_{bar}+m_{H}}$ is the reduced mass of
the bound state picture. (In the effective theory, $m_{bar}$ corresponds to
the mass of the nucleon, while in the bound state picture this light baryon
mass is taken to be the lowest odd parity resonance state of the nucleon. To
this order in the expansion---$\mathcal{O}(\lambda)$---the relative
differences in these light baryon masses are NLO effects.) The notation
$\mathcal{H}_{\lambda^{n}}$ indicates the piece of the Hamiltonian whose
contribution to the low-lying states is of order $\lambda^{n}$. The term
$\mathcal{H}_{\lambda^{\frac{1}{2}}}$ is referred to as leading order since
the terms $\mathcal{H}_{\lambda^{-1}}$ and $\mathcal{H}_{\lambda^{0}}$ are
constants and do not affect the dynamics; $\mathcal{H}_{\lambda}$ is the NLO contribution.

At $\mathcal{O}(\lambda)$ in the effective expansion, the excitation energy of
the heavy baryon is expressed in terms of two unknown phenomenological
parameters $\kappa$ and $\alpha$,
\begin{equation}
\Delta m_{\Lambda_{Q}}=\sqrt{\frac{\kappa}{\mu_{Q}}}+\frac{\alpha}{4!}\frac
{5}{\mu_{Q}\kappa}+\mathcal{O}(\lambda^{\frac{3}{2}}),
\label{ExcitationEnergyQ}%
\end{equation}
where the second term is obtained using perturbation theory. Note that in
nature, the first excited state of the heavy baryon is actually a doublet,
which in the single-channel models is taken to be the \textit{spin-averaged}%
\ \textit{mass }(\textit{i.e., }$m_{\overline{\Lambda}_{Q}^{\ast}}\equiv
\frac{1}{3}m_{\Lambda_{Q1}}+\frac{2}{3}m_{\Lambda_{Q1}^{\ast}}$, where
$m_{\Lambda_{Q1}}$ is the mass of the excited state of the doublet wit total
angular momentum $J=0$, and $m_{\Lambda_{Q1}^{\ast}}$ is the mass
corresponding to the state with $J=1$). On the other hand, for coupled-channel
models, these doublet of excited states are treated distinctly.

From the effective theory, the dominant semi-leptonic form factors for
transitions between $\Lambda_{b}$ and $\Lambda_{c}$ were also derived in ref.
\cite{hb3}. Up to NLO, these observables are completely determined in terms of
the parameters $\kappa$ and $\alpha$; there are no new parameters associated
with the currents between these states at this order. The transitions between
the ground states of $\Lambda_{b}$ and $\Lambda_{c}$ involve matrix elements
of the form,
\begin{equation}
\langle\Lambda_{c}(\vec{v}^{\prime})|\bar{c}\gamma^{\mu}(1-\gamma
_{5})b|\Lambda_{b}(\vec{v})\rangle=\bar{u}_{c}(\vec{v}^{\prime})\left(
\Gamma_{V}-\Gamma_{A}\right)  u_{b}(\vec{v})\, \label{MC}%
\end{equation}
where,
\begin{equation}
\Gamma_{V}=f_{1}\gamma^{\mu}+if_{2}\sigma^{\mu\nu}q_{\nu}+f_{3}q^{\mu}\,,
\label{J}%
\end{equation}

\begin{equation}
\Gamma_{A}=\left(  g_{1}\gamma^{\mu}+ig_{2}\sigma^{\mu\nu}q_{\nu}+g_{3}q^{\mu
}\right)  \gamma_{5}\,,
\end{equation}
where $q=m_{\Lambda_{c}}v^{\prime}-m_{\Lambda_{b}}v$ is the momentum transfer
and $u_{b}(\vec{v}^{\prime})$, $u_{c}(\vec{v})$ are the Dirac spinors
corresponding to $\Lambda_{c}$ and $\Lambda_{b}$ baryons.

While there exist in total six different form factors associated with the
current operator in eq. (\ref{MC}), in the combined expansion at NLO only two
form factors are dominant for ground-to-ground semi-leptonic decays,
\begin{align}
\left\langle \Lambda_{c}(\vec{v}^{\prime})|\overline{c}\gamma^{0}b|\Lambda
_{b}(\vec{v})\right\rangle  &  =F_{1}\overline{u}_{c}(\vec{v}^{\prime}%
)\gamma^{0}u_{b}(\vec{v})(1+\mathcal{O}(\lambda^{\frac{3}{2}}),\\
\left\langle \Lambda_{c}(\vec{v}^{\prime})|\overline{c}\gamma^{i}\gamma
_{5}b|\Lambda_{b}(\vec{v})\right\rangle  &  =G_{1}\overline{u}_{c}(\vec
{v}^{\prime})\gamma^{0}\gamma_{5}u_{b}(\vec{v})(1+\mathcal{O}(\lambda
^{\frac{3}{2}}),\nonumber
\end{align}
where $F_{1}$ and $G_{1}$ are respectively the scalar and vector form factors
for ground-to-ground transitions, and $\vec{v}$ and $\vec{v}^{\prime}$ are,
respectively, the initial and final baryon velocities. These expressions are
valid only for small velocities (\textit{i.e., }of order $\lambda^{\frac{3}%
{4}}$.) As shown in ref. \cite{hb3}, since $F_{1}$ and $G_{1}$ are equal up to
corrections of order $\lambda$ in the combined expansion, these
ground-to-ground transitions can be expressed in terms of a single form
factor. At LO, the form factor is given by:%

\begin{align}
\Theta &  =F_{1}={\frac{2\sqrt{2}\,\mu_{b}^{\frac{3}{8}}\mu_{c}^{\frac{3}{8}}%
}{(\sqrt{\mu_{b}}+\sqrt{\mu_{c}})^{\frac{3}{2}}}}\label{FF}\\
&  \exp\left(  -{\frac{m_{bar}|\delta\vec{v}|^{2}}{2(\sqrt{\kappa\mu_{b}%
}+\sqrt{\kappa\mu_{c}})}}\right)  \left(  1+\mathcal{O}(\lambda)\right)
\,,\nonumber
\end{align}
where $|\delta\vec{v}|\sim\lambda^{\frac{3}{4}}$. Expressed as a function of
the velocity transfer $|\delta\vec{v}|$, $\Theta$ has an essential singularity
in the combined limit; it vanishes faster than any power of $\lambda$.
However, it can be re-expressed as a smooth function of a new dimensionless
kinematic variable $z$ where,
\begin{equation}
z\equiv{\frac{m_{bar}\sqrt{2(w-1)}}{(\sqrt{\mu_{b}}+\sqrt{\mu_{c}})^{\frac
{1}{2}}}}\,\text{.} \label{zw}%
\end{equation}
Accordingly, we will express all observables in terms of $z$ instead of the
velocity transfer parameter $w=v\cdot v^{\prime}$ (where $v$ and $v^{\prime}$
are respectively the initial and final state baryon velocities) as is
generally done in the heavy quark expansion. Of course, there is no difference
in the information content whether the function is written in terms of $w$ or
$z$. At NLO, the semi-leptonic form factor for ground-to-ground transitions is
given by,%
\begin{equation}%
\begin{tabular}
[c]{l}%
$\Theta(z)={\frac{2\sqrt{2}\,\mu_{b}^{\frac{3}{8}}\mu_{c}^{\frac{3}{8}}%
}{(\sqrt{\mu_{b}}+\sqrt{\mu_{c}})^{\frac{3}{2}}}}\exp\left(  \frac{-z^{2}%
}{2\sqrt{\kappa}}\right)  $\\
$\left(  1+\frac{\alpha}{4!\kappa^{\frac{5}{2}}(\sqrt{\mu_{b}}+\sqrt{\mu_{c}%
})}\left(  \frac{45\kappa(\sqrt{\mu_{b}}-\sqrt{\mu_{c}})^{2}}{16\kappa
^{\frac{3}{2}}\sqrt{\mu_{b}\mu_{c}}}+{5\,z^{2}\sqrt{\kappa}}-{\frac{1}{4}%
}z^{4}\right)  \right)  \left(  1+\mathcal{O}(\lambda^{\frac{3}{2}})\right)
.$%
\end{tabular}
\ \label{Theta}%
\end{equation}

These expressions for the form factors are functions, and thus contain an
infinite amount of information. Of particular interest are the values of
$\Theta(z)$ evaluated at zero recoil and the curvature of $\Theta(z)$ at
$z=0$; by symmetry the slope vanishes at $z=0$. At NLO, $\Theta(z)$ evaluated
at zero recoil is given by,
\begin{align}
\Theta_{0}  &  \equiv\Theta(z=0)={\frac{2\sqrt{2}\,\,\mu_{b}^{\frac{3}{8}}%
\mu_{c}^{\frac{3}{8}}}{(\sqrt{\mu_{b}}+\sqrt{\mu_{c}})^{\frac{3}{2}}}}\left(
1+{\frac{\alpha}{4!}}{\frac{45(\sqrt{\mu_{b}}-\sqrt{\mu_{c}})^{2}}%
{16\kappa^{\frac{3}{2}}\sqrt{\mu_{b}\mu_{c}}(\sqrt{\mu_{b}}+\sqrt{\mu_{c}})}%
}\right) \label{Thetazero}\\
&  \left(  1+\mathcal{O}(\lambda^{\frac{3}{2}})\right)  \,.\nonumber
\end{align}

The curvature of $\Theta(z)$ evaluated at zero recoil is given by,
\begin{align}
\rho &  \equiv{\frac{\partial^{2}\Theta}{\partial z^{2}}}(z=0)=-{\frac
{2\sqrt{2}\,\,\mu_{b}^{3/8}\mu_{c}^{3/8}}{\sqrt{\kappa}(\sqrt{\mu_{b}}%
+\sqrt{\mu_{c}})^{3/2}}}\left(  1+{\frac{\alpha}{4!}}{\frac{45(\sqrt{\mu_{b}%
}-\sqrt{\mu_{c}})^{2}-160\sqrt{\mu_{b}\mu_{c}}}{16\kappa^{3/2}\sqrt{\mu_{b}%
\mu_{c}}(\sqrt{\mu_{b}}+\sqrt{\mu_{c}})}}\right) \label{rho}\\
&  \left(  1+\mathcal{O}(\lambda^{\frac{3}{2}})\right)  \,.\nonumber
\end{align}
\qquad

A $\Lambda_{b}$ can also undergo a semi-leptonic decay into the first excited
state of $\Lambda_{c}$. For these ground-to-first excited state transitions,
there exist two semi-leptonic decay channels which correspond to decays from
the ground state of $\Lambda_{b}$ to the doublet of the first excited states
of $\Lambda_{c}$: $\Lambda_{b}\rightarrow\Lambda_{c1}\ell\bar{\nu}$ and
$\Lambda_{b}\rightarrow\Lambda_{c1}^{\ast}\ell\bar{\nu}$. The dominant form
factor which determines the hadronic amplitudes for these two channels is
given in terms of a single form factor $\Xi(z)$ \cite{hb3,hb4},
\begin{align}
\langle\Lambda_{c1}(\vec{v}^{\prime})|\bar{c}\gamma^{\mu}(1-\gamma
_{5})b|\Lambda_{b}(\vec{v})\rangle &  =\sqrt{3}\,\Xi(z)\bar{u}_{c}(\vec
{v}^{\prime})\gamma^{\mu}(1-\gamma_{5})u_{b}(\vec{v})\left(  1+\mathcal{O}%
(\lambda)\right)  \,,\\
\langle\Lambda_{c1}^{\ast}(\vec{v}^{\prime})|\bar{c}\gamma^{\mu}(1-\gamma
_{5})b|\Lambda_{b}(\vec{v})\rangle &  =\Xi(z)\,\bar{u}_{c\nu}(\vec{v}^{\prime
})(\sigma^{\nu\mu}\gamma_{5}-g^{\mu\nu})u_{b}(\vec{v})\left(  1+\mathcal{O}%
(\lambda)\right)  \,,\nonumber
\end{align}
where the Rarita-Schwinger spinors are normalized according to $\bar{u}_{\nu
}(\vec{v},s)u^{\nu}(\vec{v},s)=-1$. For $z$ of order unity (i.e.\textit{, for
velocity transfers of order $\lambda^{\frac{3}{4}}$),} the form factor
$\Xi(z)$ at NLO is,%
\begin{equation}%
\begin{tabular}
[c]{l}%
$\Xi\,(z)={\frac{4\,z\,\mu_{b}^{\frac{3}{8}}\mu_{c}^{\frac{5}{8}}}%
{\kappa^{\frac{1}{4}}(\sqrt{\mu_{b}}+\sqrt{\mu_{c}})^{2}}}\exp\left(
-{\frac{z^{2}}{2\sqrt{\kappa}}}\right)  $\\
$\left(  1+{\frac{\alpha}{4!\kappa^{\frac{5}{2}}(\sqrt{\mu_{b}}+\sqrt{\mu_{c}%
})}}\left(  {\frac{(105\mu_{b}-230\sqrt{\mu_{b}\mu_{c}}+45\mu_{c})\kappa
}{16\sqrt{\mu_{b}\mu_{c}}}}+{\frac{13}{2}}\,\kappa^{1/2}z^{2}-{\frac{1}{4}%
}\,z^{4}\right)  \right)  \left(  1+\mathcal{O}(\lambda^{\frac{3}{2}})\right)
\,.$%
\end{tabular}
\ \label{Ksi}%
\end{equation}
While the form factor $\Xi(z)$ vanishes at zero recoil, the slope of $\Xi(z)$
evaluated at zero recoil does not,
\begin{align}
\sigma &  \equiv{\frac{\partial\Xi}{\partial z}}\,(z=0)={\frac{4\,\,\mu
_{b}^{\frac{3}{8}}\mu_{c}^{\frac{5}{8}}}{\kappa^{\frac{1}{4}}(\sqrt{\mu_{b}%
}+\sqrt{\mu_{c}})^{2}}}\label{sigma}\\
&  \left(  1+{\frac{\alpha}{4!}}{\frac{105\mu_{b}-230\sqrt{\mu_{b}\mu_{c}%
}+45\mu_{c}}{16\kappa^{\frac{3}{2}}\sqrt{\mu_{b}\mu_{c}}(\sqrt{\mu_{b}}%
+\sqrt{\mu_{c}})}}\right)  \left(  1+\mathcal{O}(\lambda^{\frac{3}{2}%
})\right)  \,.\nonumber
\end{align}

We will compare results for these observables based on the effective theory to
NLO to predictions obtained using potential models. For the single and
coupled-channel models, we use model currents to describe transitions between
states of $\Lambda_{b}$ and $\Lambda_{c}$. The semi-leptonic decays are
obtained using model wave functions. The most general current operator in the
effective theory derived in ref. \cite{hb2} is given in terms of the relative
position operator $\overrightarrow{x}$. The general boost operator can be
expressed in terms of the velocity transfer $\delta\overrightarrow{v}$ as,
\begin{equation}
B(\delta\vec{v})=\exp\left[  -im_{bar}\delta\vec{v}\cdot\overrightarrow
{x}\right]  \label{CurrentOP}%
\end{equation}
where $m_{bar}$ is the mass of the lowest $N^{\ast}$ state, and
$\overrightarrow{x}$ is the relative position between $N^{\ast}(1535)$ and the
heavy meson. In the effective theory, $m_{bar}$ in the boost operator
corresponds to the nucleon mass.

The total radiative decay rates of excited $\Lambda_{c}$ and $\Lambda_{b}$
baryons were also calculated in the effective theory. The total decay rates in
the charm and bottom sectors were derived up to NLO in ref. \cite{hb3} for a
single-channel transition where the doublet of first excited states are
considered as one:
\begin{equation}
\Gamma(\Lambda_{c1}\rightarrow\Lambda_{c}\,\gamma)={\frac{1}{6}}e^{2}%
\kappa{\left(  \frac{m_{\bar{D}}-m_{bar}}{m_{\bar{D}}m_{bar}}\right)  ^{2}%
}\left(  1-{\frac{\alpha}{4!}}{\frac{5}{\sqrt{\kappa^{3}\mu_{c}}}}\right)
\left(  1+\mathcal{O}(\lambda)\right)  \,, \label{Gammac}%
\end{equation}%
\begin{equation}
\Gamma(\Lambda_{b1}\rightarrow\Lambda_{b}\,\gamma)=\frac{1}{6}e^{2}%
\kappa\left(  \frac{m_{\overline{B}}+m_{bar}}{m_{\overline{B}}m_{bar}}\right)
^{2}\left(  1-{\frac{\alpha}{4!}}{\frac{5}{\sqrt{\kappa^{3}\mu_{b}}}}\right)
\left(  1+\mathcal{O}(\lambda)\right)  \,, \label{Gammab}%
\end{equation}
where $e$ is the electromagnetic coupling constant ($e^{2}\approx\frac{1}%
{137}$).

The computation of the model-independent results for the electromagnetic
decays has an important subtlety. The charges of the quarks are not
unambiguously defined. At first sight, it may seem obvious to use the physical
charge assignments for the $N_{c}=3$ world of $\frac{2}{3}$ and -$\frac{1}{3}%
$. However, this presents a problem. Recall that to order $\lambda$, there are
two complementary and equivalent pictures of the collective dynamics: one can
view the heavy baryon as a heavy quark oscillating against the brown-muck or
alternatively as a heavy meson oscillating against a light baryon.
Dynamically, this makes little difference since, for example, the mass of the
heavy quark is not so different from the mass of the heavy meson. Consider,
however, the charges. The charge of a c-quark is $\frac{2}{3}$ but the charge
of a D-meson is either 1 or zero depending on the isospin projection.
Similarly, the charge of the recoiling brown-muck will be -$\frac{2}{3}$,
while the charge of the recoiling light baryon is either -1 or zero. Thus,
there exists an order unity differences between the two pictures. Formally,
this issue can be resolved by considering the charges as given in the large
$N_{c}$ limit. This is fully consistent with the derivation of the effective
theory which is based on large $N_{c}$ arguments. Assigning charges so that
the usual baryons have charge of unity (for the proton and $\Lambda_{c}$) and
zero (for the neutron and $\Lambda_{b}$) requires that the quark charges be,
\begin{align}
e_{u}\,  &  =\,e_{c}\,=e_{t}\,=\,\frac{1}{2}+\frac{1}{{N_{c}}}\nonumber\\
e_{d}\,  &  =\,e_{s}\,=e_{b}\,=-\,\frac{1}{2}+\frac{1}{{N_{c}}}%
\end{align}
It is easy to see that, as $N_{c}\rightarrow\infty$, using these charge
assignments, the two dynamical pictures agree up to $\frac{1}{N_{c}}$
corrections. Thus, at a formal level, there is no difficulty in predicting the
observables of eqs.~(\ref{Gammac}) and~(\ref{Gammab}). However, there is a
problem phenomenologically. The very large difference between the two
descriptions at $N_{c}=3$ suggests that for the physical world one expects
very large $\frac{1}{N_{c}}$ corrections; thus we do not expect the effective
theory to be accurate for these observables and will not report on them in
this paper.

At leading order in the effective theory, all observables depend on a single
parameter, $\kappa$, and all other observables may be predicted once one
observable is fixed. In practice, we already know the splitting between the
ground state of $\Lambda_{c}$ and its first excited state. Hence, we can use
this splitting to predict all remaining observables to this order. These
observables were derived in \cite{hb4}, and are listed in Table [I]. At NLO in
the expansion, the parameters $\kappa$ and $\alpha$ can be eliminated to give
model-independent predictions for the physical observables of $\Lambda_{c}$
and $\Lambda_{b}$ baryons. At $\mathcal{O}(\lambda)$, each observable can be
expressed in terms of two additional observables \cite{hb4}. Thus, once two of
these observables are measured experimentally, the other observable can easily
be predicted. At present, only one observable of $\Lambda_{Q}$ baryons has
been measured---the excitation energy of $\Lambda_{c}$. Accordingly, it is
useful to express all of the observables in terms of this known splitting and
one of the other observables; when one measurement gets made, we will then be
able to predict the other observables to this order. These relations were
derived in \cite{hb4} and are reported in Table [II]. The predictions of Table
[II] require some explanation. We re-scale the constants $\kappa$ and $\alpha$
as well as the observables using the typical momentum scale of the collective
degrees of freedom. This re-scaling allows us to parameterize the size of the
NLO corrections in a simple way. It is defined as: $\Lambda\equiv(\mu
_{c}\Delta m_{\Lambda_{c}}^{2})^{\frac{1}{3}}\thickapprox$ 410 MeV.%

\section{Single-Channel Potential Models}

\label{SCSection}In this section, we study the phenomenology of isoscalar
heavy baryons based on the bound state picture of ref. \cite{bst3}. We first
examine the spectroscopy of $\Lambda_{c}$ and $\Lambda_{b}$ baryons using a
variety of single-channel potentials. The bound states of these systems
correspond to solutions of the Schr\"{o}dinger equation in which no
distinction is made between the H and H$^{\ast}$ mesons. Our failure to
distinguish between these two heavy mesons is reasonable in the context of the
effective theory at NLO since the splitting between these states only occurs
at NNLO. Thus, the single-channel models do not test all possible ways the
effective theory can break down. In practice, however, what these models probe
are effects due to anharmonicity in the collective motion. Recall that, in the
effective theory at NLO, only the leading harmonic correction was included; in
these models, however, harmonic effects are included to all orders. We study
coupled-channel models later; these studies give insights into the role played
by the splitting of the H and H$^{\ast}$ mesons.

For single-channel potentials, the Schr\"{o}dinger equation is given by,
\begin{equation}
u^{\prime\prime}(r)+2\mu\left(  E-V(r)-\frac{\ell(\ell+1)}{2\mu r^{2}}\right)
u(r)=0, \label{sceq.}%
\end{equation}
where the reduced wave function is given by $u(r)=rR(r)$, $\ell$ represents
the orbital angular momentum between the light baryon and heavy meson, $V(r)$
is the single-channel potential between the light baryon and heavy meson in
the bound state picture, and $E$ is the bound state energy. The reduced mass
$\mu$ of the bound state picture is given in terms of the mass of the light
baryon and the spin-averaged mass of the heavy meson (H and H$^{\ast}$),
\begin{equation}
\mu=\frac{m_{bar}m_{\overline{H}}}{m_{bar}+m_{\overline{H}}}.
\end{equation}
In accordance with the bound state picture, the light baryon mass $m_{bar}$ is
taken to be the mass of the lowest odd parity resonance state of the nucleon
namely N$^{\ast}(1535).$ (In the quark model, this may be interpreted as a
state where one quark is in an L=1 excited state.) In the charm sector,
$\mu\thickapprox863$ MeV, while $\mu\thickapprox1191$ MeV in the bottom
sector. The N$^{\ast}(1535)$ has the following quantum numbers: $I(J^{P}
)=\frac{1}{2}\left(  \frac{1}{2}^{-}\right)  $, where $I$ \ is the total
isospin of N$^{\ast}$, $J$ is the total spin, and $\ P$ is its parity. Thus,
by coupling N$^{\ast}$ to the heavy meson H (with $I(J^{P})=$ $\frac{1}
{2}\left(  0^{-}\right)  $), we obtain consistent quantum numbers with those
of $\Lambda_{Q}$.

Equation (\ref{sceq.}) is solved by standard means. By fitting the ground and
first excited state energy eigenvalue solutions of eq. (\ref{sceq.}) to the
experimental values for the ground and first excited states of $\Lambda_{c}$,
we can, in principle, predict additional excited states below threshold.
However, since we don't believe the higher-lying states---these states lie
beyond the range where we believe the expansion to be valid---we only consider
the excitation energy between the ground and first orbitally excited states in
both the charm and bottom sectors. We use model wave functions (\textit{i.e.},
solutions to eq. (\ref{sceq.})) as well as model currents given in eq.
(\ref{CurrentOP}) to calculate the physical observables of eqs.
(\ref{ExcitationEnergyQ}), (\ref{Thetazero}), (\ref{rho}), and (\ref{sigma}).

While there exist infinitely many potentials that can be used to describe the
low-lying bound states of heavy baryons, we explore a few; we hope these are
enough to reach qualitative conclusions on the convergence of the effective
expansion. As we will see, certain models work better than others. We choose
simple potential models which describe the low-lying states of heavy baryons
based on the effective theory. These potential models are simple in form in
that they have a minimum at the origin, and die off at sufficiently large
distances. For both the single and coupled-channel models, the integration
limit (\textit{i.e.},where the potential becomes weak,) lies between 3 and 4
$fm$. We will study the following potential models :
\[%
\begin{tabular}
[c]{l}%
$V(r)=-c_{0}e^{-(\frac{r}{a_{0}})^{2}},$\\
$V(r)=-c_{0}\frac{e^{-(\frac{r}{a_{0}})^{2}}}{1+(\frac{r}{a_{0}})^{2}},$\\
$V(r)=-\frac{c_{0}}{2}(e^{-(\frac{r}{a_{0}})^{2}}+e^{-(\frac{r}{a_{0}})^{4}%
}),$\\
$V(r)=-\frac{c_{0}}{2}\frac{\left(  e^{-(\frac{r}{a_{0}})^{2}}+e^{-(\frac
{r}{a_{0}})^{4}}\right)  }{1+(\frac{r}{a_{0}})^{2}+(\frac{r}{a_{0}})^{4}},$\\
$V(r)=-\frac{c_{0}}{2}(e^{-(\frac{r}{a_{0}})^{2}}+e^{-(2\frac{r}{a_{0}})^{2}%
}),$\\
$V(r)=-c_{0}e^{-\left(  1+(\frac{r}{a_{0}})^{2}\right)  ^{\frac{1}{2}}+1}.$%
\end{tabular}
\ \ \ \ \
\]
In general, the parameters of these models are fixed so that the ground and
first excited state energies are fitted to the mass of $\Lambda_{c}$ and the
spin-averaged mass of the $\Lambda_{c}$ doublet.

\section{Predictions Based on Single-Channel Models}

Using the various single-channel models, we obtain consistent predictions for
the physical observables of $\Lambda_{c}$ and $\Lambda_{b}$ baryons. For these
models, the first excited odd parity resonance mass of the nucleon is used in
place of the nucleon mass, which is used in the effective theory, for parity
reasons. This choice presents no ambiguity since the difference in mass
between these two nucleon states occurs at NNLO in the effective expansion.
For these models, the predicted excitation energy for $\Lambda_{b}$ differs
from the LO prediction based on the effective theory by at most 20 MeV. The
prediction for $\Theta_{0}$ (the form factor for ground-to-ground transitions
evaluated at zero recoil) is consistent with the LO prediction of unity
obtained in the combined expansion and in HQET. The model prediction for
$\rho$ (the curvature of the form factor for ground-to-ground transitions
evaluated at zero recoil) is approximately -1.50$\times$10$^{-4}$
MeV$^{-\frac{3}{2}}$. This prediction differs from the LO value (where
$\rho\thickapprox-1.50\times10^{-4}$ MeV$^{-\frac{3}{2}}$) by 25\%. The model
prediction for $\sigma$ (the slope of the form factor for ground-to-first
excited state transitions evaluated at zero recoil) is approximately 0.018
MeV$^{-\frac{3}{4}}$. This prediction differs by 60\% from the LO prediction
based on the effective theory, where $\sigma\thickapprox0.011$MeV$^{-\frac
{3}{4}}$.

\section{Coupled-Channel Potential Models}

\label{CCPotModels}In this section, we describe the formalism for obtaining
coupled-channel models which can be used to calculate the physical observables
of $\Lambda_{c}$ and $\Lambda_{b}$ baryons. These potential models are of
particular interest because they allow us to test another possible source of
error in the predictions at NLO in the effective theory---namely, the effects
due to the splitting of the heavy mesons. The coupled-channel models requires
an excited state of the heavy baryon. In some models (see \cite{bst2}), the
bound state picture is based on a soliton-heavy meson model in which the
soliton incorporates the internal excitations of the brown-muck. In our bound
state picture, however, these internal excitations correspond to excited
states of the nucleon. Here we will consider an explicit excited state of the
light baryon which lies above the lowest-lying odd parity N$^{\ast}$(1535)
state. We choose the N$^{\ast\ast}$(2090) state for two reasons: it has the
right quantum numbers as those of N$^{\ast}$(1535) (\textit{i.e.},
$I(J^{P})\rightarrow\frac{1}{2}(\frac{1}{2}^{-})$ in accordance with the usual
spectroscopic notation where $I$ is the particle isospin, $J$ is the total
spin of the particle, and $P$ is its parity). In addition, its energy lies
approximately 550 MeV above the N$^{\ast}$(1535) state. Thus, by choosing this
particular excited nucleon resonance state, we maintain an energy splitting
characteristic of typical internal excitations of the brown-muck. Moreover,
when we couple the N$^{\ast}$(1535) and the N$^{\ast\ast}$(2090) states to the
heavy mesons H and H$^{\ast}$ in the bound state picture, we obtain states
with consistent quantum numbers as those of the heavy baryon.

For coupled-channel models, we need to map the bound state picture onto the
heavy quark and brown-muck picture in order to obtain two pictures with
consistent quantum numbers as those of the heavy baryon $\Lambda_{Q}.$ We set
up the mapping in such a way that, in the limit where $\lambda=0$, we obtain
two identical displaced copies of the excitation energy. This displaced energy
is due to the splitting between the states of the light baryon (\textit{i.e.}%
,N$^{\ast}$(1535) and N$^{\ast\ast}$(2090)). When the splitting between the H
and H$^{\ast}$ states goes to zero, we retrieve the single-channel results.
While this mapping strategy is not the most general approach to define
coupled-channel models, it is appropriate in our case because it retrieves the
single-channel model-independent results in the $\lambda\ \rightarrow0$ limit.
Thus, in this limit, the heavy quark--brown-muck and bound state pictures
become equivalent.

In the heavy quark and brown-muck picture, the heavy baryon ground state can
be written as some collective wave function times
\begin{equation}
\left\vert \frac{1}{2},\frac{1}{2}\right\rangle _{Q}\left\vert
0,0\right\rangle _{bm},
\end{equation}
where $Q$ denotes the heavy quark state and $bm$ denotes the brown-muck state.
First, in order to obtain a hadronic picture with quantum numbers consistent
with the heavy quark and brown-muck\ picture (which has the brown-muck in an
isosinglet state with total spin zero), we need to remove a $q\overline{q}$
(quark and anti-quark) pair with total spin zero from the vacuum. This action
leaves the brown-muck in a total spin zero state. We then couple $\overline
{q}$ to the heavy quark spin to make a meson with total spin $J=0$ or $1$.
Finally, we couple $q$ to the brown-muck to make an excited nucleon (in this
case, an N$^{\ast}$ or an N$^{\ast\ast}$) with total spin $\frac{1}{2}$. By
using appropriate {Clebsch-Gordan coefficients}, the total wave function is
expressed in terms of a superposition of H and H$^{\ast}$ and N$^{\ast}$ and
N$^{\ast\ast}$state corresponding to the coupling of the heavy meson to
excited nucleons is given by,

\begin{align}
&  \frac{1}{\sqrt{2}}\left(  \left\vert \frac{1}{2},\frac{1}{2}\right\rangle
_{Q}\left\vert \frac{1}{2},-\frac{1}{2}\right\rangle _{\overline{q}%
}+\left\vert \frac{1}{2},-\frac{1}{2}\right\rangle _{Q}\left\vert \frac{1}%
{2},\frac{1}{2}\right\rangle _{\overline{q}}\right)  \left(  \left\vert
\frac{1}{2},\frac{1}{2}\right\rangle _{N^{\ast}}+\left\vert \frac{1}{2}%
,\frac{1}{2}\right\rangle _{N^{\ast\ast}}\right)  +\label{HQBMS}\\
&  \frac{1}{\sqrt{2}}\left(  \left\vert \frac{1}{2},\frac{1}{2}\right\rangle
_{Q}\left\vert \frac{1}{2},-\frac{1}{2}\right\rangle _{\overline{q}%
}-\left\vert \frac{1}{2},-\frac{1}{2}\right\rangle _{Q}\left\vert \frac{1}%
{2},\frac{1}{2}\right\rangle _{\overline{q}}\right)  \left(  \left\vert
\frac{1}{2},\frac{1}{2}\right\rangle _{N^{\ast}}+\left\vert \frac{1}{2}%
,\frac{1}{2}\right\rangle _{N^{\ast\ast}}\right) \nonumber
\end{align}
times the appropriate collective wave function. In eq. (\ref{HQBMS}), the
first element in the bra-kets denotes the total spin of the particle, and the
second element is the projection. The symmetric wave function given by
$\frac{1}{\sqrt{2}}\left(  |\frac{1}{2},\frac{1}{2}\rangle_{Q}|\frac{1}%
{2},-\frac{1}{2}\rangle_{\overline{q}}+|\frac{1}{2},-\frac{1}{2}\rangle
_{Q}|\frac{1}{2},\frac{1}{2}\rangle_{\overline{q}}\right)  $ corresponds to
the heavy meson state H, while the anti-symmetric state $\frac{1}{\sqrt{2}%
}\left(  |\frac{1}{2},\frac{1}{2}\rangle_{Q}|\frac{1}{2},-\frac{1}{2}%
\rangle_{\overline{q}}-|\frac{1}{2},-\frac{1}{2}\rangle_{Q}|\frac{1}{2}%
,\frac{1}{2}\rangle_{\overline{q}}\right)  $ is the state of H$^{\ast}$.

In the bound state picture, the heavy mesons H and H$^{\ast}$ correspond to
the D and D$^{\ast}$ mesons and to the B and B$^{\ast}$ mesons, respectively.
These states have the following quantum numbers: $D^{0}\rightarrow\frac{1}%
{2}(0^{-}),$ $D^{\ast}\rightarrow\frac{1}{2}(1^{-}),$ $B^{0}\rightarrow
\frac{1}{2}(0^{-}),$ and $B^{\ast}\rightarrow\frac{1}{2}(1^{-})$. The states
of the bound state picture, which include the heavy meson states and nucleon
resonance states, can all be expressed in terms of states of well-defined
angular momentum,
\begin{equation}
|s,\ell,J,M_{J}\rangle=\sum_{m_{s}m_{\ell}}\left\langle j_{1}j_{2,}%
,m_{1},m_{2}\right\vert \left\vert J,M;j_{1}j_{2}\right\rangle |s,m_{s}%
\rangle|\ell,m_{\ell}\rangle,
\end{equation}
where $s$\ is the total spin of the light baryon and heavy meson ($\,H$ or
$\,H^{\ast}$) and $m_{s}$ represents its projection, $\ell$ is the relative
orbital angular momentum between the light baryon and the heavy meson ($\,$H
or $\,$H$^{\ast}$) and $m_{\ell}$ is its projection, $J=|s+\ell|,|s+\ell
-1|....|s-\ell|$ is the total spin, and $M=m_{s}+m_{\ell}$ represents the
projection of the total spin. Note that, as discussed earlier, there is no
spin-orbit contribution in this description. Using this notation, the ground
state (\textit{i.e.}, the $\ell=0$ state) of the bound state\ picture can be
expressed in terms of a particular linear combination of the states H and
H$^{\ast}$and N$^{\ast}$and N$^{\ast\ast}$ as follows:
\begin{align}
&  \alpha\left(  \left\vert \frac{1}{2},\frac{1}{2}\right\rangle _{N^{\ast}%
}+\left\vert \frac{1}{2},\frac{1}{2}\right\rangle _{N^{\ast\ast}}\right)
\left\vert 0,0\right\rangle _{H}+\label{GSWF}\\
&  \beta\left(  \sqrt{\frac{2}{3}}\left(  \left\vert \frac{1}{2},-\frac{1}%
{2}\right\rangle _{N^{\ast}}+\left\vert \frac{1}{2},-\frac{1}{2}\right\rangle
_{N^{\ast\ast}}\right)  \left\vert 1,1\right\rangle _{H^{\ast}}-\sqrt{\frac
{1}{3}}\left(  \left\vert \frac{1}{2},\frac{1}{2}\right\rangle _{N^{\ast}%
}+\left\vert \frac{1}{2},\frac{1}{2}\right\rangle _{N^{\ast\ast}}\right)
\left\vert 1,0\right\rangle _{H^{\ast}}\right) \nonumber
\end{align}
times the collective wave function; $\alpha$ and $\beta$ are coefficients
specifying the correct superposition with the obvious constraint that
$\alpha^{2}+\beta^{2}=1$. The bra-kets in eq. (\ref{GSWF}) are expressed,
respectively, in terms of the total spin of the particle and its projection.
The coefficients $\alpha$ and $\beta$ can easily be obtained by matching the
ground state of the bound state picture in eq. (\ref{GSWF}) to the ground
state of the heavy quark and brown-muck picture in eq. (\ref{HQBMS}). This
yields,
\begin{equation}
\alpha=\frac{1}{2}\text{ and }\beta=\frac{\sqrt{3}}{2}. \label{coeff}%
\end{equation}

Because we take the heavy meson states H and H$^{\ast}$ to be nondegenerate
for coupled-channel potentials, we need to include these two orthogonal states
in our mapping procedure. We should note that at order $\lambda$ in the
effective expansion, the orbital contribution decouples from the dynamics so
that the orthogonal states of the bound state picture are given by,
\begin{align}
&  \alpha\left(  \left\vert \frac{1}{2},\frac{1}{2}\right\rangle _{N^{\ast}%
}+\left\vert \frac{1}{2},\frac{1}{2}\right\rangle _{N^{\ast\ast}}\right)
\left\vert 0,0\right\rangle _{H}+\label{OS1}\\
&  \beta\left(  \sqrt{\frac{2}{3}}\left(  \left\vert \frac{1}{2},-\frac{1}%
{2}\right\rangle _{N^{\ast}}+\left\vert \frac{1}{2},-\frac{1}{2}\right\rangle
_{N^{\ast\ast}}\right)  \left\vert 1,1\right\rangle _{H^{\ast}}-\sqrt{\frac
{1}{3}}\left(  \left\vert \frac{1}{2},\frac{1}{2}\right\rangle _{N^{\ast}%
}+\left\vert \frac{1}{2},\frac{1}{2}\right\rangle _{N^{\ast\ast}}\right)
\left\vert 1,0\right\rangle _{H^{\ast}}\right) \nonumber
\end{align}
and,%
\begin{align}
&  -\beta\left(  \left\vert \frac{1}{2},\frac{1}{2}\right\rangle _{N^{\ast}%
}+\left\vert \frac{1}{2},\frac{1}{2}\right\rangle _{N^{\ast\ast}}\right)
\left\vert 0,0\right\rangle _{H}+\label{OS2}\\
&  \alpha\left(  \sqrt{\frac{2}{3}}\left(  \left\vert \frac{1}{2},-\frac{1}%
{2}\right\rangle _{N^{\ast}}+\left\vert \frac{1}{2},-\frac{1}{2}\right\rangle
_{N^{\ast\ast}}\right)  \left\vert 1,1\right\rangle _{H^{\ast}}-\sqrt{\frac
{1}{3}}\left(  \left\vert \frac{1}{2},\frac{1}{2}\right\rangle _{N^{\ast}%
}+\left\vert \frac{1}{2},\frac{1}{2}\right\rangle _{N^{\ast\ast}}\right)
\left\vert 1,0\right\rangle _{H^{\ast}}\right)  .\nonumber
\end{align}
times the collective wave function. Next, we transform the Hamiltonian from
the heavy quark and brown-muck basis (where at least at low orders in the
expansion we know its form) to a hadronic basis where the states are given in
terms of the coefficients of the orthogonal basis states of the bound state
picture in eqs. (\ref{OS1}) and (\ref{OS2}). Note that, as discussed above, in
the heavy quark and brown-muck basis, we include two copies of the
dynamics---one based on the ground state of the brown-muck and the other based
on an excited state of the brown-muck. We will treat these two states as
having the same collective dynamics (apart from small reduced mass effects
which contribute beyond NLO in the effective theory). Moreover, these states
are displaced from each other by the excitation energy of the brown-muck. The
reason for doing this should be clear---once we have made this transformation
to the hadronic basis we are then in a position to add perturbations due to
the splitting of the two heavy meson states. The transformation is
\begin{equation}
\mathcal{H}^{\prime}=\Phi^{-1}\mathcal{H}\Phi
\end{equation}
where $\Phi$ contains the coefficients of the orthogonal states (see eq.
(\ref{coeff})) of the transformation from the heavy quark and brown-muck
picture to the bound state picture:
\begin{equation}
\Phi=\left(
\begin{array}
[c]{cc}%
\alpha & \beta\\
-\beta & \alpha
\end{array}
\right)  .
\end{equation}
The effective Hamiltonian, $\mathcal{H}$, can be written as follows:
\[
\left(
\begin{array}
[c]{cc}%
\mathcal{H}_{1} & 0\\
0 & \mathcal{H}_{2}%
\end{array}
\right)
\]
where,%

\begin{equation}
\mathcal{H}_{1}=\mathcal{H}_{S}\ ,\text{ and } \label{H1}%
\end{equation}%
\begin{equation}
\mathcal{H}_{2}=\mathcal{H}_{S}+\Delta+(m_{H}-m_{\overline{H}}), \label{H2}%
\end{equation}
where $\mathcal{H}_{S}$ denotes the single-channel Hamiltonian, $\Delta$ is
the mass splitting between the N$^{\ast}$(1535) and N$^{\ast\ast}$(2090), and
$\overline{H}$ is the spin-averaged mass of the heavy meson. Thus, the bound
state Hamiltonian in the coupled channel can be written as,
\begin{align}
\mathcal{H}^{\prime}  &  =\left(
\begin{array}
[c]{cc}%
\alpha & -\beta\\
\beta & \alpha
\end{array}
\right)  \left(
\begin{array}
[c]{cc}%
\mathcal{H}_{1} & 0\\
0 & \mathcal{H}_{2}%
\end{array}
\right)  \left(
\begin{array}
[c]{cc}%
\alpha & \beta\\
-\beta & \alpha
\end{array}
\right) \label{Hp}\\
&  =\left(
\begin{array}
[c]{cc}%
\alpha^{2}\mathcal{H}_{1}+\beta^{2}\mathcal{H}_{2} & \alpha\beta
(\mathcal{H}_{1}-\mathcal{H}_{2})\\
\alpha\beta(\mathcal{H}_{1}-\mathcal{H}_{2}) & \beta^{2}\mathcal{H}_{1}%
+\alpha^{2}\mathcal{H}_{2}%
\end{array}
\right)  .\nonumber
\end{align}
Plugging eqs. (\ref{H1}) and (\ref{H2}) into eq. (\ref{Hp}), the Hamiltonian
for the bound state picture reduces to:
\begin{equation}
\mathcal{H}^{\prime}=\left(
\begin{array}
[c]{cc}%
\mathcal{H}_{s}^{bar\,H}+\beta^{2}\Delta+(m_{H}-m_{\overline{H}}) &
-\alpha\beta\Delta\\
-\alpha\beta\Delta & \mathcal{H}_{s}^{bar\,H^{\ast}}+\alpha^{2}\Delta
+(m_{H^{\ast}}-m_{\overline{H}})
\end{array}
\right)
\end{equation}

We can now write down the corresponding coupled Schr\"{o}dinger equations,
\begin{align}
&  u^{\prime\prime}(r)+2\mu_{barH}\left(  E-V_{barH}(r)-\frac{\ell(\ell
+1)}{2\mu_{barH}r^{2}}-\frac{3}{4}\Delta-(m_{H}-m_{\overline{H}})\right)
u(r)\label{CC1}\\
+2\alpha\beta\Delta\mu_{barH}w(r)  &  =0,\nonumber
\end{align}

\begin{align}
&  w^{\prime\prime}(r)+2\mu_{barH^{\ast}}\left(  E-V_{bar\,H^{\ast}}%
(r)-\frac{\ell(\ell+1)}{2\mu_{barH^{\ast}}r^{2}}-\frac{1}{4}\Delta
-(m_{H^{\ast}}-m_{\overline{H}})\right)  w(r)\label{CC2}\\
+2\alpha\beta\Delta\mu_{barH^{\ast}}u(r)  &  =0,\nonumber
\end{align}
where $u(r)$ is the bound state reduced wave function, with reduced mass
$\mu_{barH}$, of the light baryon and H, and $w(r)$ is the bound state reduced
wave function, with reduced mass $\mu_{barH^{\ast}}$, of the light baryon and
H$^{\ast}$. In eqs. (\ref{CC1}) and (\ref{CC2}), the light baryon corresponds
to the lowest odd parity nucleon resonance state N$^{\ast}$(1535).

There exists an ambiguity in the reduced masses in the coupled-channel
formalism. It is unclear as to what masses one should use since the orthogonal
states of the bound state picture involve linear combinations of H and
H$^{\ast}$ as well as N$^{\ast}$ and N$^{\ast\ast}$ (as can be seen in eqs.
(\ref{OS1}) and (\ref{OS2})). For concreteness, we express the reduced masses
in the coupled channels in terms of bound states of H with N$^{\ast}$(1535)
and H$^{\ast}$ with N$^{\ast}$(1535) because the difference between these
reduced masses is small, in accordance with the effective theory
(\textit{i.e., }the corrections are NNLO effects.) Again, we remind the reader
that the error in the reduced masses in the coupled channel should be of the
same order as that in the effective theory since the effective theory is
supposed to describe the \textquotedblleft real world\textquotedblright. These
masses are distinguished for no other reason than to take into account mass
splitting effects and to fit the form of the potentials.

To determine the collective wave functions of heavy baryons, we solve eqs.
(\ref{CC1}) and (\ref{CC2}) by imposing standard boundary conditions such
that:
\begin{align}
u_{r\rightarrow0}(r)  &  \rightarrow r^{\ell+1}\rightarrow0\text{ as
}r\rightarrow0\text{ and }\label{bcc}\\
u_{r\rightarrow\infty}(r)  &  \rightarrow\sqrt{r}(J_{k}%
(z)+Y_{k}(z))\rightarrow0\text{ as }\nonumber\\
z  &  =\left(  \sqrt{2\mu_{N^{\ast}H}E}r\text{ or }\sqrt{2\mu_{N^{\ast}%
H^{\ast}}Er}\right)  \rightarrow\infty,\nonumber
\end{align}
\qquad where $J_{k}(z)$ is the Bessel function of the first kind, and
$Y_{k}(z)$ is the Bessel function of the second kind. Here $k=\frac{1}%
{2}(1+2\ell)$, where $\ell$ is the relative orbital angular momentum between
the excited nucleon and the heavy meson. The values of the potential depth
$c_{0}$ and the length parameter $a_{0}$ are fitted by requiring that the
predicted energies of the ground and first excited states of the bound state
picture be matched to the experimental ground and first doublet of excited
state energies of $\Lambda_{c}$,
\begin{align}
m_{\Lambda_{c}}  &  =\Lambda_{c}-(m_{bar}+m_{\overline{D}})\thickapprox
-1221\text{ MeV,}\\
m_{\Lambda_{c}^{^{\prime}}}  &  =\Lambda_{c}^{\prime}-(m_{bar}+m_{\overline
{D}})\thickapprox-892\text{ MeV,}\nonumber
\end{align}
where $m_{bar}$ is the mass the N$^{\ast}$(1535) state, $m_{\overline{D}}$ is
the spin-averaged \textit{ }mass of the D and D$^{\ast}$ mesons, and where
$m_{\Lambda_{c}^{^{\prime}}}$ is the spin-averaged mass of the doublet of
first excited states of $\Lambda_{c}$. The ground and first doublet of excited
states of heavy baryons are, in general, written as:
\begin{align}
|\Lambda_{Q}\rangle &  \equiv|\Lambda_{Q};0,{\frac{1}{2}},J_{z}\rangle
\,\sim\Lambda_{c},\,\,\,\Lambda_{b}\,,\nonumber\\
|\Lambda_{Q1}\rangle &  \equiv|\Lambda_{Q};1,{\frac{1}{2}},J_{z}\rangle
\,\sim\Lambda_{c}(2593),\,\,\,\Lambda_{b}(?)\,,\nonumber\\
|\Lambda_{Q1}^{\ast}\rangle &  \equiv|\Lambda_{Q};1,{\frac{3}{2}},J_{z}%
\rangle\,\sim\Lambda_{c}(2625),\,\,\,\Lambda_{b}(?)\,. \label{states}%
\end{align}

While the reduced masses of $\Lambda_{c}$ and $\Lambda_{b}$ are not equal in
the combined limit where $\lambda\rightarrow0$, these two sectors are modeled
to have the same potential forms. This is valid up to NLO in the effective
theory. Thus, higher-order corrections of the difference between the two
potential forms will be ignored. This is not done for any deep reason---just
for simplicity. Again, it is rigorously correct in the $\lambda\rightarrow0$
limit. This allows us to substitute the values of $\kappa$ and $\alpha$
extracted from the charm sector using coupled-channel models to make
quantitative predictions for the observables in both the charm and bottom
sectors. Once again, we remind the reader that we have chosen simple
potentials with two free parameters---the potential depth $c_{0}$ and a length
parameter $a_{0}$, characteristic of hadronic scales. These potentials
describe the low-lying states of heavy baryons based on the effective theory.
The parameters $\kappa$ and $\alpha$ in the effective Hamiltonian of eq.
(\ref{EffHam}) are calculated directly from the potential energy once $c_{0}$
and $a_{0}$ are determined for a given potential. At $\mathcal{O}(\lambda)$ in
the effective expansion, the total potential energy of the heavy baryon is
given by,
\begin{equation}
V(\overrightarrow{x})=-c_{0}+m_{bar}+m\overline{_{H}}+\kappa\frac
{\overrightarrow{x}^{2}}{2!}+\alpha\frac{\overrightarrow{x}^{4}}%
{4!}+\mathcal{O}(\lambda^{\frac{3}{2}}).
\end{equation}
The value of $\kappa$ in the effective Hamiltonian is obtained in the bound
state picture by determining the value of $\frac{d^{2}V(r)}{dr^{2}}|_{r=0}$.
Similarly, $\alpha$ can be measured by evaluating $\frac{d^{4}V(r)}{dr^{4}%
}|_{r=0}$. Using these parameters extracted from the potential models, the
observables obtained in the effective theory at NLO in eqs.
(\ref{ExcitationEnergyQ}), (\ref{Thetazero}), (\ref{rho}), and (\ref{sigma})
can be calculated. In addition to the Hamiltonian, the currents are taken to
be identical in structure to those obtained in the effective theory with no
new parameters (valid only up to order $\lambda$ in the expansion.)
Higher-order terms are neglected for simplicity. Although the excitation
energy of $\Lambda_{b}$ has not yet been measured experimentally, it can be
predicted by substituting the values of $\kappa$ and $\alpha$ obtained from
the $\Lambda_{c}$ fit into the differential equation for the bound state
picture of $\Lambda_{b}$ (see eqs. (\ref{CC1}) and (\ref{CC2})) with reduced
mass given by,
\begin{equation}
\ \mu_{b}=\frac{m_{bar}m\overline{_{B}}}{m_{bar}+m\overline{_{B}}},
\end{equation}
where $m_{\overline{B}}$ is the spin-averaged mass of the B-meson.

\section{Predictions Based on Coupled-Channel Models}

\label{CCModpredictions}We use the coupled-channel model derived in section
\ref{CCPotModels} to compute the physical observables discussed previously.
Tables [\ref{T9}-\ref{T14}] are arranged as follows. The second column lists
the model predictions for the physical observables of $\Lambda_{c}$ and
$\Lambda_{b}$ baryons. The third column contains the model-independent
predictions based on the effective theory at LO, and the last two columns give
model-independent predictions (with parameters extracted from models) based on
the effective theory to NLO.

Our results based on the coupled-channel models show that the predicted
excitation energy in the bottom sector differs at most by 40 MeV from the LO
prediction based on the effective theory of 300 MeV (a 13\% error which lies
within the expected error at this order.) We also notice agreement in the
model predictions with typical errors for the semi-leptonic decays. The model
prediction for $\Theta_{0}$ is consistent with the LO prediction of unity
obtained from the effective theory. The model prediction for $\rho$ is
approximately -0.90$\times$10$^{-4}$ MeV$^{-\frac{3}{2}}$---a 25\% difference
from the LO prediction, where $\rho\thickapprox$ $-1.20\times10^{-4}$
MeV$^{-\frac{3}{2}}$. The coupled-channel model prediction for $\sigma$ is
approximately 0.0080 MeV$^{-\frac{3}{4}}$---a 27\% difference from the LO
prediction, where $\sigma\thickapprox0.011$ MeV$^{-\frac{3}{4}}$.%

\section{Discussion}

\label{ConclusionSec}In this paper, we tested the model independent
predictions for the observables of isoscalar heavy baryons (which include
$\Lambda_{c}$ and $\Lambda_{b}$ baryons) obtained from an effective theory
based on a combined heavy quark and large $N_{c}$ expansion (derived in ref.
\cite{hb4}) against model predictions based on the bound state picture of
heavy baryons. The model predictions (obtained using \textquotedblleft
reasonable\textquotedblright\ potentials) are compared with LO and NLO
predictions obtained from the effective theory. The model-independent
expressions for the observables of $\Lambda_{c}$ and $\Lambda_{b}$ baryons
were derived to order $\lambda$ in the effective theory, where $\lambda$ is
the natural expansion parameter. By fitting the excitation energy of
$\Lambda_{c}$ to experiment for both the single and coupled-channel models, we
were able to predict the excitation energy of $\Lambda_{b}$ as well as the
semi-leptonic decays of $\Lambda_{c}$ and $\Lambda_{b}$ baryons.

By calculating the properties of $\Lambda_{c}$ and $\Lambda_{b}$ baryons using
potential models, we explored the usefulness of the effective theory in the
combined expansion. In principle, we want to test the convergence of the
effective theory. Naively, one would expect the predictions based on the
effective theory at LO to be rather crude since the expansion is in powers of
$\lambda^{\frac{1}{2}}\thickapprox\frac{1}{\sqrt{3}}$---a 60\% error. At NLO,
we might expect the error in the predictions for the physical observables to
be approximately 30\%.

While the potential models we have chosen do not necessarily describe the
``real world,'' they do describe the low-lying
states of heavy baryons in a manner consistent with the effective theory.
Using various single-channel models, consistent predictions for the low-lying
excitation energy of $\Lambda_{b}$ were obtained (\textit{i.e., }$\Delta
m_{\Lambda_{b}}\thickapprox288$ MeV). This predicted energy differs from the
LO prediction of 300 MeV by approximately 4\%. For coupled-channel models, on
the other hand, the predicted excitation energy of $\Lambda_{b}$ differs by at
most 13\% from the LO prediction. Moreover, when we compare the single-channel
model predictions for the excitation energy of $\Lambda_{b}$ to the
coupled-channel model predictions, we notice that these models differ from
each other by at most 10\%.

Based on our calculations, the single and coupled-channel model predictions
for the semi-leptonic decays of $\Lambda_{c}$ and $\Lambda_{b}$ baryons are in
close agreement with the NLO model-independent predictions (with parameters
extracted from the models). The model prediction for $\Theta_{0}$ (the
semi-leptonic form factor for ground-to-ground transitions evaluated at zero
recoil given in eq. (\ref{Thetazero}) to NLO) is consistent for both the
single and coupled-channel models with the LO prediction of unity obtained in
the combined expansion and in HQET.

Comparing the single and coupled-channel model predictions for $\rho$ (the
curvature of the semi-leptonic form factor for ground-to-ground transitions
evaluated at zero recoil given in eq. (\ref{rho})) with the NLO
model-independent predictions for $\rho$, we notice close agreement with a
degree of accuracy expected from an NLO expansion. For the single-channel
models, the prediction for $\rho$ (where $\rho\thickapprox-1.50\times10^{-4}$
MeV$^{-\frac{3}{2}}$) differ from the model-independent predictions (expressed
in terms of the excitation energy of $\Lambda_{c}$ and $\sigma$) (where
$\rho\thickapprox-1.26\times10^{-4}$ MeV$^{-\frac{3}{2}}$) by approximately
20\%. This prediction agrees with the LO prediction of -1.20$\times$10$^{-4}$
MeV$^{-\frac{3}{2}}$ to within 25\%---an error which lies well within the
expected error. Moreover, for the coupled-channel models, the prediction for
$\rho$ is approximately -0.90$\times$10$^{-4}$ MeV$^{-\frac{3}{2}}$. This
prediction differs from the LO prediction based on the effective expansion
(where $\rho\thickapprox$-1.20$\times$10$^{-4}$ MeV$^{-\frac{3}{2}}$) by 25\%.
The NLO model-independent predictions for $\rho$ also agree for all models
(\textit{i.e., }$\rho\thickapprox-1.18\times10^{-4}$ MeV$^{-\frac{3}{2}}$) to
within the expected accuracy. This prediction lies within 2\% of the LO
prediction for $\rho$. Thus, based on these predictions, the expansion for
\textit{ }$\rho$ appears to be useful at NLO.

For the single-channel models, $\sigma\thickapprox0.018$ MeV$^{-\frac{3}{4}}%
$---a 60\% difference from the LO prediction based on the effective expansion
(where $\sigma\thickapprox0.011$ MeV$^{-\frac{3}{4}}$). The NLO
model-independent prediction for $\sigma$ (expressed in terms of the
excitation energy of $\Lambda_{c}$ and $\rho$) is approximately 0.013
MeV$^{-\frac{3}{4}}$. This prediction differs from the LO prediction based on
the effective theory by 18\%, and agrees with the model to within
28\%.\ Similar agreement for the $\sigma$ observables is seen for the
coupled-channel models . The prediction for $\sigma$ is approximately 0.0080
MeV$^{-\frac{3}{4}}$, which differs from the LO prediction of 0.011
MeV$^{-\frac{3}{4}}$ by 27\%. Again, this error lies within the expected
error. The prediction for $\sigma$ is also seen to agree with the
model-independent predictions for $\sigma$ (where $\sigma\thickapprox$ 0.0087
MeV$^{-\frac{3}{4}}$) to the expected level of accuracy. This prediction
differs from the LO prediction by about 21\%. Again, based on these
predictions, the expansion for $\sigma$ also appears to be useful at NLO.

By comparing the single and coupled-channel model predictions for $\Theta_{0}%
$, $\rho$, and $\sigma$ in Tables [III-XIV] to the respective
model-independent predictions (with parameters based on model results), we
notice that the expansion seems to work rather well for these observables.
Thus, it is plausible that these model independent predictions might be useful
in describing the phenomenology of isoscalar heavy baryons in the real world.
However, there is a caveat which should be made. While the model-independent
predictions based on the $\lambda$ expansion appear to work for our models to
the expected accuracy, the difference in results between the single and
coupled-channel models gives a cause for concern. Note that if the logic
underlying our expansion is correct, the difference between these two types of
models should appear at NNLO or higher. To the extent that the expansion at
NLO gives reliable results, we would expect the difference between the
predictions of the two types of models to be modest---of order of the
uncertainties and thus typically of the same order as the discrepancies
between the model-independent NLO predictions and the results of the models.
Instead, the disagreements between the predictions of the two models are in
fact much larger. For example, one sees that the single-channel model
predictions for $\rho$ are typically -1.50$\times$10$^{-4}$ MeV$^{-\frac{3}%
{4}}$, while the coupled channel models had $\rho\thickapprox$-0.090$\times
$10$^{-4}$ MeV$^{-\frac{3}{4}}$. Thus, the single-channel results are nearly
1.7 times the coupled-channel results. Moreover, the discrepancy for $\sigma$
is even larger; the single-channel model predictions are more than double
those of the coupled channel models.

The discrepancy between the single and coupled-channel model predictions seems
to suggest that the parameters in the models and (by inference presumably
those of the real world) are such that we may well be beyond the region of
useful convergence of the expansion, at least for these observables. How can
we reconcile the apparent success of the model-independent predictions based
on the expansion with the differences between the two classes of models? A
reasonable conjecture is that the expansion is rather marginal ({\it i.e.,}
whether the expansion is useful at NLO depends on details.) We see that the
expansion is more robust for some observables, or combinations of observables,
than for others. This will happen if certain combinations of observables lead
to effects of higher order terms largely canceling, while other combinations
lead to these effects adding more coherently. Presumably, the combinations of
observables in our model-independent predictions relating $\rho$ and $\sigma$
and the splitting of the $\Lambda_{c}$ states are robust in this sense at
least in the context of these models. This is not too surprising. In fact, one
might well expect correlations between the effects of higher-order terms on
the form factor for decays to the ground state of $\Lambda_{c}$ and its
excited states. It is not unreasonable to hope that a similar robustness may
apply for the real world so that the $\lambda$ expansion may be useful in
making predictions. It will be interesting to see if this is the case \ when
measurements of the semi-leptonic decay form-factors become available.

\begin{acknowledgments}
Support of this research was funded by the U.S. Department of Energy under
grant no. DE-FG02-93ER-40762. Z. A. Baccouche acknowledges the support of the
Southeastern Universities Research Association for its SURA/Jefferson Lab
fellowship, and Fermi Laboratory for its fellowship which she received during
her graduate studies at the University of Maryland. She also wishes to thank
James Kelly for his help with numerical algorithms. The authors acknowledge
Boris Gelman for useful discussions. This work is from a dissertation to be
submitted to the Graduate School , University of Maryland, by Z. A. Baccouche
in partial fulfillment of the requirements for the Ph.D. degree in Physics.
\end{acknowledgments}

\newif\ifabfull\abfulltrue

\clearpage
\begin{table}[tbh] \centering
%

\begin{tabular}
[t]{|l|l|l|l|l|}\hline
$\Delta m_{\Lambda_{c}}$ (MeV) & $\Delta m_{\Lambda_{b}}$ (MeV) & $\Theta_{0}$
& $\rho$ (MeV$^{-\frac{3}{2}}$) & $\sigma$ (MeV$^{-\frac{3}{4}}$)\\\hline
$330\,$ & $300$ & $0.99$ & $-1.20\times10^{-4}$ & $0.011$\\\hline
\end{tabular}%
\caption{These are the model-independent predictions based on the effective
theory at LO. Here $\Delta m_{\Lambda_{c}}%
$ is the spin-averge excitation energy
of $\Lambda_{c}$ obtained from experiement, $\Delta m_{\Lambda_{b}}$ is the
predicted excitation energy of $\Lambda_{b}$ to LO
(i.e, $\Delta m_{\Lambda_{b}}\thickapprox\sqrt{\frac{\kappa}{\mu_{b}}}$ where
$\kappa\thickapprox(410$ MeV$)^{3}$ to LO and where $\mu_{b}$ is the
reduced mass of $\Lambda_{b}$). $\Theta_{0}%
$ is the form factor evaluated at zero
reoil ($z=0$) for transitions between the ground state of $\Lambda_{b}$ to the
ground state of $\Lambda_{c}$ (see eq.(13)). The second derivative of the form
factor
for ground-to-ground transitons evaluated at zero recoil is given in eq.(14) by
$\rho$, and $\sigma
$ is the derivative of the form factor evaluated at zero recoil for
transitions between the ground state of $\Lambda_{b}%
$ and first excited state of
$\Lambda_{c}$ (see eq.(17)). \label{T1}}%
\end{table}%
%

\begin{table}[tbh] \centering
\begin{tabular}
[c]{|c|c|c|}\hline
Predictions & $\Delta m_{\Lambda_{c}}$ and $\rho$ & $\Delta m_{\Lambda_{c}}$
and $\sigma$\\\hline
\multicolumn{1}{|l|}{$\Delta m_{\Lambda_{b}}\,\Lambda^{-1}$} &
\multicolumn{1}{|l|}{$1.29+0.57\,(\rho\Lambda^{3/2})$} &
\multicolumn{1}{|l|}{$0.77-0.05\,(\sigma\Lambda^{3/4})$}\\\hline
\multicolumn{1}{|l|}{$\Theta_{0}$} & \multicolumn{1}{|l|}{$0.95-0.05\,(\rho
\Lambda^{3/2})$} & \multicolumn{1}{|l|}{$0.99+0.004\,(\sigma\Lambda^{3/4})$%
}\\\hline
\multicolumn{1}{|l|}{$\rho\,\Lambda^{3/2}$} & \multicolumn{1}{|l|}{} &
\multicolumn{1}{|l|}{$-0.92-0.08\,(\sigma\Lambda^{3/4})$}\\\hline
\multicolumn{1}{|l|}{$\sigma\,\Lambda^{3/4}$} &
\multicolumn{1}{|l|}{$0.25-0.72\,(\rho\Lambda^{3/2})$} &
\multicolumn{1}{|l|}{}\\\hline
\end{tabular}%
\caption
{These are the model independent predictions based on the effective theory
for the observables
$\Delta m_{\Lambda_{b}}$, $\Theta_{0}$, $\rho$, and $\sigma$ given in
eqs.(5),(13),(14), and (17) at NLO. These observables are rescaled in terms of
$\Lambda$ (where $\Lambda\equiv(\mu_{c}(\Delta m_{\Lambda_{c}})^{2}%
)^{(1/3)}=410$ MeV)
and are expressed at NLO in terms of the known excitation energy
of $\Lambda_{c}$ and one additional observable.\label{T2}}%
\end{table}%

\begin{table}[tbh] \centering
\begin{tabular}
[c]{|c|c|c|c|c|}\hline
& Model Predictions & Fitted LO & $\Delta m_{\Lambda_{c}}$ and $\rho$ &
$\Delta m_{\Lambda_{c}}$ and $\sigma$\\\hline
$\Delta m_{\Lambda_{b}}$ (MeV) & 288 & 300 & 236 & 282\\\hline
$\Theta_{0}$ & 0.99 & 0.99 & 1.01 & 0.99\\\hline
$\rho$ (MeV$^{-\frac{3}{2}}$) & -1.51$\times$10$^{-4}$ & -1.20$\times$%
10$^{-4}$ &  & -1.27$\times$10$^{-4}$\\\hline
$\sigma$ (MeV$^{-\frac{3}{4}}$) & 0.018 & 0.011 & 0.013 & \\\hline
\end{tabular}%
\caption{This table gives the predictions for the excitation energy of
$\Lambda_{b}$ and the semileptonic form factors given to NLO in
eqs. (5), (13), (14) and (17).
The second column gives predictions for the observables based on the single
channel model potential
$V(r)=-c_{0}\exp[-(\frac{r}{a_{0}}^{2})]$.
The third column gives the leading-order predictions for the observables based on the
effective theory. The last two columns give model predictions for the
observables based on the model-independent predictions obtained from the effective
theory at NLO (see Table [II]).\label{T3}}%
\end{table}%
%

\begin{table}[tbh] \centering
\begin{tabular}
[c]{|l|l|l|l|l|}\hline
& Model Predictions & Fitted LO & $\Delta m_{\Lambda_{c}}$ and $\rho$ &
$\Delta m_{\Lambda_{c}}$ and $\sigma$\\\hline
$\Delta m_{\Lambda_{b}}$ (MeV) & 289 & 300 & 237 & 281\\\hline
$\Theta_{0}$ & 0.99 & 0.99 & 1.012 & 0.99\\\hline
$\rho$ (MeV$^{-\frac{3}{2}}$) & -1.50$\times$10$^{-4}$ & -1.20$\times$%
10$^{-4}$ &  & -1.27$\times$10$^{-4}$\\\hline
$\sigma$ (MeV$^{-\frac{3}{4}}$) & 0.019 & 0.011 & 0.013 & \\\hline
\end{tabular}%
\caption{This table gives the predictions for the excitation energy of
$\Lambda_{b}$ and the semileptonic form factors given to NLO in eqs. (5),
(13), (14) and (17).
The second column gives predictions for the observables based on the single-channel model potential
$V(r)=-c_{0}\frac{\exp\left[ -\left( \frac{r}{a_{0}}\right) ^{2}\right
] }{1+\left( \frac{r}{a_{0}}\right) ^{2}}$.
The third column gives the leading-order predictions for the observables based on the
effective theory. The last two columns give model predictions for the
observables based on the model-independent predictions obtained from the effective
theory at NLO (see Table [II]).\label{T4}}%
\end{table}%
%

\begin{table}[tbh] \centering
\begin{tabular}
[c]{|l|l|l|l|l|}\hline
& Model Predictions & Fitted LO & $\Delta m_{\Lambda_{c}}$ and $\rho$ &
$\Delta m_{\Lambda_{c}}$ and $\sigma$\\\hline
$\Delta m_{\Lambda_{b}}$ (MeV) & 280 & 300 & 231 & 284\\\hline
$\Theta_{0}$ & 0.99 & 0.99 & 1.014 & 0.99\\\hline
$\rho$ (MeV$^{-\frac{3}{2}}$) & -1.54$\times$10$^{-4}$ & -1.20$\times$%
10$^{-4}$ &  & -1.26$\times$10$^{-4}$\\\hline
$\sigma$ (MeV$^{-\frac{3}{4}}$) & 0.017 & 0.011 & 0.013 & \\\hline
\end{tabular}%
\caption{This table gives the predictions for the excitation energy of
$\Lambda_{b}$ and the semileptonic form factors given to NLO in  eqs. (5),
(13), (14) and (17).
The second column gives predictions for the observables based on the single-channel model potential
$V(r)=-\frac{c_{0}}{2}(\exp\left[ -\left( \frac{r}{a_{0}}\right) ^{2}%
\right] +\exp\left[ -\left( \frac{r}{a_{0}}\right) ^{4}\right] )$.
The third column gives the leading-order predictions for the observables based on the
effective theory. The last two columns give model predictions for the
observables based on the model-independent predictions obtained from the effective
theory at NLO (see Table [II]).\label{T5}}%
\end{table}%
%

\begin{table}[tbh] \centering
\begin{tabular}
[c]{|l|l|l|l|l|}\hline
& Model Predictions & Fitted LO & $\Delta m_{\Lambda_{c}}$ and $\rho$ &
$\Delta m_{\Lambda_{c}}$ and $\sigma$\\\hline
$\Delta m_{\Lambda_{b}}$ (MeV) & 285 & 300 & 234 & 283\\\hline
$\Theta_{0}$ & 0.99 & 0.99 & 1.013 & 0.99\\\hline
$\rho$ (MeV$^{-\frac{3}{2}}$) & -1.52$\times$10$^{-4}$ & -1.20$\times$%
10$^{-4}$ &  & -1.26$\times$10$^{-4}$\\\hline
$\sigma$ (MeV$^{-\frac{3}{4}}$) & 0.018 & 0.011 & 0.013 & \\\hline
\end{tabular}%
\caption{This table gives the predictions for the excitation energy of
$\Lambda_{b}$ and the semileptonic form factors given to NLO in eqs. (5),
(13), (14) and (17).
The second column gives predictions for the observables based on the single-channel model potential
$V(r)=-\frac{c_{0}}{2}\frac{(\exp\left[ -\left( \frac{r}{a_{0}}\right
) ^{2}\right] +\exp\left[ -\left( \frac{r}{a_{0}}\right) ^{4}\right
] )}{1+\left( \frac{r}{a_{0}}\right) ^{2}+\left( \frac{r}{a_{0}}\right) ^{4}%
}$.
The third column gives the leading-order predictions for the observables based on the
effective theory. The last two columns gives model predictions for the
observables based on the model-independent predictions obtained from the effective
theory at NLO (see Table [II]).\label{T6}}%
\end{table}%
%

\begin{table}[tbh] \centering
\begin{tabular}
[c]{|l|l|l|l|l|}\hline
& Model Predictions & Fitted LO & $\Delta m_{\Lambda_{c}}$ and $\rho$ &
$\Delta m_{\Lambda_{c}}$ and $\sigma$\\\hline
$\Delta m_{\Lambda_{b}}$ (MeV) & 292 & 300 & 238 & 280\\\hline
$\Theta_{0}$ & 0.99 & 0.99 & 1.012 & 0.99\\\hline
$\rho$ (MeV$^{-\frac{3}{2}}$) & -1.50$\times$10$^{-4}$ & -1.20$\times$%
10$^{-4}$ &  & -1.28$\times$10$^{-4}$\\\hline
$\sigma$ (MeV$^{-\frac{3}{4}}$) & 0.019 & 0.011 & 0.013 & \\\hline
\end{tabular}%
\caption{This table gives the predictions for the excitation energy of
$\Lambda_{b}$ and the semileptonic form factors given to NLO in eqs. (5),
(13), (14) and (17).
The second column gives predictions for the observables based on the single-channel model potential
$V(r)=-\frac{c_{0}}{2}(\exp\left[ -\left( \frac{r}{a_{0}}\right) ^{2}%
\right] +\exp\left[ -\left( 2\frac{r}{a_{0}}\right) ^{2}\right] )$.
The third column gives the leading-order predictions for the observables based on the
effective theory. The last two columns give model predictions for the
observables based on the model-independent predictions obtained from the effective
theory at NLO (see Table [II]).\label{T7}}%
\end{table}%
%

\begin{table}[tbh] \centering
\begin{tabular}
[c]{|l|l|l|l|l|}\hline
& Model Predictions & Fitted LO & $\Delta m_{\Lambda_{c}}$ and $\rho$ &
$\Delta m_{\Lambda_{c}}$ and $\sigma$\\\hline
$\Delta m_{\Lambda_{b}}$ (MeV) & 293 & 300 & 239 & 279\\\hline
$\Theta_{0}$ & 0.99 & 0.99 & 0.99 & 0.99\\\hline
$\rho$ (MeV$^{-\frac{3}{2}}$) & -1.50$\times$10$^{-4}$ & -1.20$\times$%
10$^{-4}$ &  & -1.28$\times$10$^{-4}$\\\hline
$\sigma$ (MeV$^{-\frac{3}{4}}$) & 0.020 & 0.011 & 0.013 & \\\hline
\end{tabular}%
\caption{This table gives the predictions for the excitation energy of
$\Lambda_{b}$ and the semileptonic form factors given to NLO in eqs. (5),
(13), (14) and (17).
The second column gives predictions for the observables based on the single-channel model potential
$V(r)=-\frac{c_{0}}{2}(\exp[-[1+\left( \frac{r}{a_{0}}\right) ^{2}%
]^{1/2}+1])$.
The third column give the leading-order predictions for the observables based on the
effective theory. The last two columns gives model predictions for the
observables based on the model-independent predictions obtained from the effective
theory at NLO (see Table [II]).\label{T8}}%
\end{table}%

\begin{table}[tbh] \centering
\begin{tabular}
[c]{|l|l|l|l|l|}\hline
& Model Predictions & Fitted LO & $\Delta m_{\Lambda_{c}}$ and $\rho$ &
$\Delta m_{\Lambda_{c}}$ and $\sigma$\\\hline
$\Delta m_{\Lambda_{b}}$ (MeV) & 306 & 300 & 353 & 300\\\hline
$\Theta_{0}$ & 0.98 & 0.99 & 0.99 & 0.99\\\hline
$\rho$ (MeV$^{-\frac{3}{2}}$) & -0.91$\times$10$^{-4}$ & -1.20$\times$%
10$^{-4}$ &  & -1.18$\times$10$^{-4}$\\\hline
$\sigma$ (MeV$^{-\frac{3}{4}}$) & 0.0082 & 0.011 & 0.0087 & \\\hline
\end{tabular}%
\caption{This table gives the predictions for the excitation energy of
$\Lambda_{b}$ and the semileptonic form factors given to NLO in eqs. (5),
(13), (14) and (17).
The second column gives predictions for the observables based on the coupled-channel model potential
$V(r)=-c_{0}\exp[-(\frac{r}{a_{0}}^{2})]$.
The third column gives the leading-order predictions for the observables based on the
effective theory. The last two columns give model predictions for the
observables based on the model-independent predictions obtained from the effective
theory at NLO (see Table [II]).\label{T9}}%
\end{table}%
%

\begin{table}[tbh] \centering
\begin{tabular}
[c]{|l|l|l|l|l|}\hline
& Model Predictions & Fitted LO & $\Delta m_{\Lambda_{c}}$ and $\rho$ &
$\Delta m_{\Lambda_{c}}$ and $\sigma$\\\hline
$\Delta m_{\Lambda_{b}}$ (MeV) & 324 & 300 & 353 & 300\\\hline
$\Theta_{0}$ & 0.97 & 0.99 & 0.99 & 0.99\\\hline
$\rho$ (MeV$^{-\frac{3}{2}}$) & -0.88$\times$10$^{-4}$ & -1.20$\times$%
10$^{-4}$ &  & -1.18$\times$10$^{-4}$\\\hline
$\sigma$ (MeV$^{-\frac{3}{4}}$) & 0.011 & 0.0076 & 0.0085 & \\\hline
\end{tabular}%
\caption{This table gives the predictions for the excitation energy of
$\Lambda_{b}$ and the semileptonic form factors given to NLO in eqs. (5),
(13), (14) and (17).
The second column gives predictions for the observables based on the coupled-channel model potential
$V(r)=-c_{0}\frac{\exp\left[ -\left( \frac{r}{a_{0}}\right) ^{2}\right
] }{1+\left( \frac{r}{a_{0}}\right) ^{2}}$.
The third column gives the leading-order predictions for the observables based on the
effective theory. The last two columns give model predictions for the
observables based on the model-independent predictions obtained from the effective
theory at NLO (see Table [II]).\label{T10}}%
\end{table}%
%

\begin{table}H] \centering
\begin{tabular}
[c]{|l|l|l|l|l|}\hline
& Model Predictions & Fitted LO & $\Delta m_{\Lambda_{c}}$ and $\rho$ &
$\Delta m_{\Lambda_{c}}$ and $\sigma$\\\hline
$\Delta m_{\Lambda_{b}}$ (MeV) & 290 & 300 & 352 & 300\\\hline
$\Theta_{0}$ & 0.96 & 0.99 & 0.99 & 0.99\\\hline
$\rho$ (MeV$^{-\frac{3}{2}}$) & -0.92$\times$10$^{-4}$ & -1.20$\times$%
10$^{-4}$ &  & -1.18$\times$10$^{-4}$\\\hline
$\sigma$ (MeV$^{-\frac{3}{4}}$) & 0.0084 & 0.011 & 0.0088 & \\\hline
\end{tabular}%
\caption{This table gives the predictions for the excitation energy of
$\Lambda_{b}$ and the semileptonic form factors given to NLO in eqs. (5),
(13), (14) and (17).
The second column gives predictions for the observables based on the coupled-channel model potential
$V(r)=-\frac{c_{0}}{2}(\exp\left[ -\left( \frac{r}{a_{0}}\right) ^{2}%
\right] +\exp\left[ -\left( \frac{r}{a_{0}}\right) ^{4}\right] )$.
The third column gives the leading-order predictions for the observables based on the
effective theory. The last two columns give model predictions for the
observables based on the model-independent predictions obtained from the effective
theory at NLO (see Table [II]).\label{T11}}%
\end{table}%

\bigskip%
\begin{table}[tbh] \centering
\begin{tabular}
[c]{|l|l|l|l|l|}\hline
& Model Predictions & Fitted LO & $\Delta m_{\Lambda_{c}}$ and $\rho$ &
$\Delta m_{\Lambda_{c}}$ and $\sigma$\\\hline
$\Delta m_{\Lambda_{b}}$ (MeV) & 290 & 300 & 351 & 301\\\hline
$\Theta_{0}$ & 0.98 & 0.99 & 0.99 & 0.99\\\hline
$\rho$ (MeV$^{-\frac{3}{2}}$) & -0.92$\times$10$^{-4}$ & -1.20$\times$%
10$^{-4}$ &  & -1.18$\times$10$^{-4}$\\\hline
$\sigma$ (MeV$^{-\frac{3}{4}}$) & 0.0081 & 0.011 & 0.0088 & \\\hline
\end{tabular}%
\caption{This table gives the predictions for the excitation energy of
$\Lambda_{b}$ and the semileptonic form factors given to NLO in eqs. (5),
(13), (14) and (17).
The second column gives predictions for the observables based on the coupled-channel model potential
$V(r)=-\frac{c_{0}}{2}\frac{(\exp\left[ -\left( \frac{r}{a_{0}}\right
) ^{2}\right] +\exp\left[ -\left( \frac{r}{a_{0}}\right) ^{4}\right
] )}{1+\left( \frac{r}{a_{0}}\right) ^{2}+\left( \frac{r}{a_{0}}\right) ^{4}%
}$.
The third column gives the leading-order predictions for the observables based on the
effective theory. The last two columns give model predictions for the
observables based on the model-independent predictions obtained from the effective
theory at NLO (see Table [II]).\label{T12}}%
\end{table}%
%

\begin{table}[tbh] \centering
\begin{tabular}
[c]{|l|l|l|l|l|}\hline
& Model Predictions & Fitted LO & $\Delta m_{\Lambda_{c}}$ and $\rho$ &
$\Delta m_{\Lambda_{c}}$ and $\sigma$\\\hline
$\Delta m_{\Lambda_{b}}$ (MeV) & 318 & 300 & 357 & 302\\\hline
$\Theta_{0}$ & 0.97 & 0.99 & 0.99 & 0.99\\\hline
$\rho$ (MeV$^{-\frac{3}{2}}$) & -0.89$\times$10$^{-4}$ & -1.20$\times$%
10$^{-4}$ &  & -1.18$\times$10$^{-4}$\\\hline
$\sigma$ (MeV$^{-\frac{3}{4}}$) & 0.0076 & 0.011 & 0.0086 & \\\hline
\end{tabular}%
\caption{This table gives the predictions for the excitation energy of
$\Lambda_{b}$ and the semileptonic form factors given to NLO in eqs. (5),
(13), (14) and (17).
The second column gives predictions for the observables based on the coupled-channel model potential
$V(r)=-\frac{c_{0}}{2}(\exp\left[ -\left( \frac{r}{a_{0}}\right) ^{2}%
\right] +\exp\left[ -\left( 2\frac{r}{a_{0}}\right) ^{2}\right] )$.
The third column gives the leading-order predictions for the observables based on the
effective theory. The last two columns give model predictions for the
observables based on the model-independent predictions obtained from the effective
theory at NLO (see Table [II]).\label{T13}}%
\end{table}%
%

\begin{table}[tbh] \centering
\begin{tabular}
[c]{|l|l|l|l|l|}\hline
& Model Predictions & Fitted LO & $\Delta m_{\Lambda_{c}}$ and $\rho$ &
$\Delta m_{\Lambda_{c}}$ and $\sigma$\\\hline
$\Delta m_{\Lambda_{b}}$ (MeV) & 261 & 300 & 360 & 302\\\hline
$\Theta_{0}$ & 0.97 & 0.99 & 0.99 & 0.99\\\hline
$\rho$ (MeV$^{-\frac{3}{2}}$) & -0.87$\times$10$^{-4}$ & -1.20$\times$%
10$^{-4}$ &  & -1.18$\times$10$^{-4}$\\\hline
$\sigma$ (MeV$^{-\frac{3}{4}}$) & 0.0076 & 0.011 & 0.0085 & \\\hline
\end{tabular}%
\caption{This table gives the predictions for the excitation energy of
$\Lambda_{b}$ and the semileptonic form factors given to NLO in eqs. (5),
(13), (14) and (17).
The second column gives predictions for the observables based on the coupled-channel model potential
$V(r)=-\frac{c_{0}}{2}(\exp[-[1+\left( \frac{r}{a_{0}}\right) ^{2}%
]^{1/2}+1])$.
The third column gives the leading-order predictions for the observables based on the
effective theory. The last two columns give model predictions for the
observables based on the model-independent predictions obtained from the effective
theory at NLO (see Table [II]).\label{T14}}%
\end{table}%

\end{document}